\documentstyle[times,pramana,epsf,floats]{ias}
\begin{document}
\mark{{A comparative study of model ingredients:....}{Sanjeev Kumar and Suneel Kumar}}
\title{A comparative study of model ingredients: fragmentation in heavy-ion collisions
using quantum molecular dynamics model}

\author{SANJEEV KUMAR and SUNEEL KUMAR$^*$}
\address{School of Physics and Material Science, Thapar University, Patiala-147004, Punjab (India)\\
$^*$suneel.kumar@thapar.edu}
\keywords{cross-section, model ingredients, quantum molecular dynamics, 
intermediate mass fragments, multifragmentation}
\pacs{25.70.Pq, 25.70.-z, 24.10.Lx}
\abstract{We aim to understand the role of NN cross-sections, equation of state 
as well as different model ingredients such as width of
Gaussian, clusterisation range and different clusterisation algorithms in
multifragmentation using quantum molecular dynamics model. We notice that
all model ingredients have sizable effect on the fragment pattern.
}
\maketitle
\section{Introduction}
The study of heavy-ion collisions at intermediate energies
$(50 \le E \le 1000~MeV/nucleon)$ provides a rich source of information for 
many rare phenomena such as multifragmentation, collective flow as well as particle 
production\cite{Puri96,Bond95}.
One can also shed light on the mechanism behind the fragmentation in highly excited
nuclear systems. In this energy region, multifragmentation appears to be a dominant de-excitation channel apart
from the other less populated channels of manifestation of liquid gas phase transition in finite nuclear systems
\cite{Puri96,Elli94,Mish98}. In the literature, multifragmentation has also been considered as 
a gateway to nuclear equation of state
\cite{Hart98,Pal98}.
Numerous investigations are cited in the literature which handle the de-excitation of
nuclear system in
multifragmentation \cite{Sing00,Mila02,Ogil91,Souz91,Tsan93,Peas94,Li06}.\\
The experimental analysis of the emission of intermediate mass fragments (IMF's), 
$(5 \le A \le A_{tot}/6)$, has
yielded several interesting observations: De Souza et al.\cite{Souz91} observed a linear
increase in the multiplicity of IMF's with incident energies
for central collisions. In this study, incident energy was varied between 35 and
110 MeV/nucleon. On the other hand,
Tsang et al. \cite{Tsan93} reported a rise and fall in the production of IMF's.
The maximal value of the IMF's shifts from nearly central to peripheral
collisions with the increase in the incident energy.
More refined results were reported by Peaslee et al. \cite{Peas94} for the reaction of
$_{36}Kr^{84}~+~_{79}Au^{197}$ for incident energy between 35 and 400 MeV/nucleon. 
Their analysis revealed that
IMF's multiplicity first increases till 100 MeV/nucleon and then decreases slowly.
These findings pose a stringent test for any theoretical model
designed for the study of multifragmentation.\\

Theoretically, multifragmentation can be studied by statistical \cite{Bond95} as well as semi-classical 
\cite{Hart98} models, respectively.
The relation between multifragmentation process and nuclear equation of state was extensively studied by several authors
with in the statistical approach for intermediate energy heavy-ion collisions
\cite{Bond95,Mish98,Pal98}. On the other hand,
semi-classical dynamical models \cite{Hart98} are very useful for 
studying the reaction from the start to final state 
where matter is fragmented and cold. In addition,
these models also give possibility to extract the information about the nuclear equation of state \cite{Li06} and
NN cross-section\cite{Puri96,Hart98,Sood09}.
The interaction among nucleons (in a heavy-ion reaction) can be studied within
the G-matrix, with its real part
representing the mean field and complex part denotes the NN cross-section 
\cite{Puri96,Hart98,Sood09}.
Note that the contribution of imaginary part of the interaction is nearly absent 
in the low energy 
process such as fusion, fission and radioactivity\cite{Gupt93}.
One often uses a parametrized form for
the real and imaginary parts of
the G-matrix. It is well accepted to use a density dependent
Skyrme type interactions for the real part of the G-matrix. However, heavy-ion dynamics depends not only on the density but also on the
entire momentum plane \cite{Sing00}. Therefore,
it is advisable to use momentum dependent interactions additionally.\\
The exact nature of NN cross-section, on the other hand, is still an open question
\cite{Sing00,Sood09}.  A large number of calculations exist in the
literature suggesting different strength and forms of the NN cross-sections \cite{Sood09,Li93}.
In a simple assumption of hard core radius of NN potential,
one has often used a constant and isotropic cross-section of 40 mb. In other calculations, 
a constant and isotropic
cross-section with magnitude between 20 and 55 mb is also used \cite{Sood09}. 
The most sensitive observable to pin down the NN cross-section is collective flow.
Recent calculations advocated its strength between 35-40 mb\cite{Sood09}. 
We shall concentrate here
on multifragmentation.
Our present study will be based on the semi-classical model, namely, quantum 
molecular dynamics (QMD). In semi-classical model, one has to face the problem of 
the stability of nuclei. Several model ingredients such as width of the Gaussian (L),
has been used as free parameter. It varies between 4.33 $fm^2$ to 8.66 $fm^2$\cite{Hart98}.
On the same time, one needs to identify the clusters with the help of clusterization 
algorithm. Different cluster recognisation algorithms can also influence the 
fragmentation.
Our present interest is to perform a comparative study of different model
ingredients alognwith different NN cross-sections and to see whether it is possible 
to pin down the strength of NN cross-sections by QMD model or not.\\
Our article is organized as follow:
we discuss the model briefly in section-II. Our results and discussions are given in section-III and we summarize
the results in section-IV. \\

\section{Quantum Molecular Dynamics (QMD) Model}
The QMD model \cite{Puri96,Hart98,Sing00,Sood09,Huan93} is a time dependent N-body theory which simulates the time evolution of heavy-ion reactions on
an event-by-event basis. It is based on the generalized variational principle. As with every variational approach,
it requires the choice of a test wave function $\Phi$. In the QMD approach, this is an N-body wave function with 6N
time-dependent parameters if nuclear system contains N nucleons.\\
The basic assumption of QMD model is that each nucleon is represented by coherent states of the form (we set $\hbar$, c = 1)
which are characterized by 6 time dependent parameters, $\vec{r_i}$ and $\vec{p_i}$, respectively.
\begin{equation}
\phi_{i}(\vec{r}, t) = \left (\frac{2}{L\pi}\right)^{3/4}\exp^{-(\vec{r}-\vec{r_{i}})^2/L}\exp^{\iota(\vec{r}
\cdot\vec{p_i}(t))},
\end{equation}
Here $L$ is the width of the Gaussian distribution. This width varies between 4.33 and
8.66 $fm^{2}$ in the literature \cite{Hart98}.
The total N-body wave function is assumed to be a direct product of the coherent states
\begin{equation}
\Phi = \Pi_{i=1}^{A_T + A_P}\phi_i(\vec{r},\vec{r_i},\vec{p_i},t)~~\cdot
\end{equation}
To calculate the time evolution of a system, we start out from the action
\begin{equation}
S = \int_{t_1}^{t_2} \pounds[\Phi, \Phi^{\ast}]dt
\end{equation}
with the Lagrange functional
\begin{equation}
\pounds = \langle~\Phi|\iota\hbar\frac{d}{dt}-H|\Phi~\rangle~~\cdot
\end{equation}
The total time derivative includes the derivation with respect to parameters. The time evolution of
parameters is obtained by the requirement that the action is stationary under the allowed variation of wave function.
This leads to a Euler-Lagrange equation for each time-dependent parameter.\\
Thus, the variational principle reduces the time evolution of N-body Schr$\ddot{o}$dinger equation to the time evolution
equations of $6(A_T + A_P)$ parameters to which a physical meaning can be attributed. The equations of motion
for the parameters $\vec{p_i}$ and $\vec{r_i}$ reads as:
\begin{equation}
\frac{d\vec{r_i}}{dt}~=~\frac{\partial< H >}{\partial \vec{p_i}}~~;~~\frac{d\vec{p_i}}{dt}~=~-\frac{\partial< H >}
{\partial \vec{r_i}}
\end{equation}
If $< H >$ has no explicit time dependence, QMD conserves the energy and momentum by construction.\\
The nuclear dynamics of the QMD model can also be translated into a semi classical scheme. The Wigner distribution
function $f_i$ of a nucleon $i^{th}$ can be easily derived from the test wave functions.
\begin{equation}
f_{i}(\vec{r},\vec{p},t)~=~\frac{1}{(\pi\hbar)^3}\exp^{-(\vec{r}-\vec{r_i}(t))^{2}\cdot2/L}\cdot\exp^{-(\vec{p}-\vec{p_i}(t))^2\cdot{L}/2\hbar^2}
\end{equation}
and the total Wigner density is the sum of those of all nucleons. Hence the expectation value of total Hamiltonian reads
\begin{eqnarray}
\langle~H~\rangle&=&\langle~T~\rangle+\langle~V~\rangle\nonumber\\
&=&\sum_{i}\frac{p_i^2}{2m_i}+
\sum_i \sum_{j > i}\int f_{i}(\vec{r},\vec{p},t)V^{\it ij}({\vec{\acute{r}},\vec{r}})\nonumber\\
& &\times f_j(\vec{\acute{r}},\vec{\acute{p}},t)d\vec{r}d\vec{r'}d\vec{p}d\vec{p'}
\end{eqnarray}
Where $V^{ij}$ is given as:
\begin{eqnarray}
V^{ij}(\vec{\acute{r}},\vec{r})&=&V^{ij}_{Skyrme}+V^{ij}_{Yukawa}+V^{ij}_{Coul}+V^{ij}_{mdi}\nonumber\\
&=&\left[t_{1}\delta(\vec{\acute{r}}-\vec{r})+t_{2}\delta(\vec{\acute{r}}-\vec{r})\rho^{\gamma-1}\left(\frac{\vec{\acute{r}}+\vec{r}}{2}\right)\right]\nonumber\\
& &+t_{3}\frac{exp(-|\vec{\acute{r}}-\vec{r}|/\mu)}{(|\vec{r'}-\vec{r}|/\mu)}~ +~\frac{Z_{i}Z_{j}e^{2}}{|\vec{\acute{r}}-\vec{r}|}\nonumber\\
& &+t_{4}\ln^2[t_{5}(\vec{\acute{p}_{i}}-\vec{p})^{2}+1]\delta(\vec{\acute{r}}-\vec{r})
\label{s1}
\end{eqnarray}
which consists of Skyrme, Yukawa, Coulomb and momentum dependent parts of the interaction and
$Z_i$, $Z_j$ are the charges of $i^{th}$ and $j^{th}$ baryons.\\
Further, potential part resulting from the convolution of the distribution function
 $\it f_{i}$ and $\it f_{j}$ with the Skyrme interactions $V_{\it Skyrme}$ reads as :
\begin{equation}
{\it V}_{Skyrme}~=~\alpha\left(\frac{\rho_{int}}{\rho_{0}}\right)+\beta\left(\frac{\rho_{int}}{\rho_{0}}\right)^{\gamma}
~~\cdot
\end{equation}
The two of the three parameters of equation of state are determined by demanding that at normal nuclear matter density,
 the binding energy should be equal to 16 MeV. The third parameter $\gamma$ is usually treated as a free parameter.
 Its value is given in term of the compressibility:
\begin{equation}
\kappa~=~9\rho^{2}\frac{\partial^{2}}{\partial\rho^{2}}\left(\frac{E}{A}\right)~~\cdot
\end{equation}
The different values of compressibility give rise to Soft and Hard equations of state.\\
The momentum dependence ($V_{mdi}$) of NN interaction, which may optionally be used in the model, is fitted to
experimental data \cite{Hart98} on the real part of the nucleon optical potential. This yields:
\begin{equation}
V_{mdi}~=~\delta\cdot ln^2 \left(\varepsilon\cdot(\Delta\vec{p})^2 + 1\right)\cdot\left(\frac{\rho_{int}}{\rho_0}\right)~~\cdot
\end{equation}
The potential part of the equation of state resulting from the convolution of distribution functions $f_{i}$
and $f_{j}$ with Skyrme and momentum dependent interactions, reads as:
\begin{eqnarray}
U&=&\alpha\left(\frac{\rho_{int}}{\rho_{0}}\right)+\beta\left(\frac{\rho_{int}}{\rho_{0}}\right)^{\gamma}\nonumber\\
 & & +\delta\cdot ln^2 \left(\varepsilon\cdot(\Delta\vec{p})^2 + 1\right)\cdot\left(\frac{\rho_{int}}{\rho_0}\right)~~\cdot
\end{eqnarray}
For the value of constants and details, reader is referred to ref. \cite{Hart98}.
The inclusion of MDI is labeled as Soft momentum dependent (SMD) and Hard momentum dependent
(HMD) equations of state. In recent studies, momentum dependent forces are found to 
be essential part of dynamics.\\
During the propagation, two nucleon can collide if they satisfy the condition
$|\vec{r_i}-\vec{r_j}|$ $\le$ $\sqrt{\sigma/\pi}$.
In this procedure,
a binary collision is blocked if the final state phase space is already occupied.
The Pauli principle of the final state reduces the free cross-section to
effective levels. Here we take different forms of
NN cross-sections [$\sigma$] to understand the influence on fragment production.
We shall here use a energy dependent cross-section due to Cugnon and also different constant cross-sections
ranging between 20 and 55 mb \cite{Hart98,Sing00}. The nature and strength of 
cross-section $\sigma$
is depicted as superscript to $\sigma$.\\

\section{Results and discussion}
\label{sec:2}
We here simulate the symmetric reactions of $_{79}Au^{197}~+~_{79}Au^{197}$ at
incident energies of 100, 400, 600 and 1000 MeV/nucleon and over complete range
of the impact parameter. The system size and asymmetry effects will be analyzed 
by further studying the reactions of
$_{10}Ne^{20}~+~_{13}Al^{27}$, $_{18}Ar^{40}~+~_{21}Sc^{45}$, $_{36}Kr^{84}~+~_{41}Nb^{93}$ and $_{54}Xe^{131}~+~_{57}La^{139}$ at energies between  $20$ and $150$ MeV/nucleon.
The clusterisation at first instance is made using the minimum spanning tree (MST) method, in
which, nucleons are bound if $R_{Clus} = |\vec{r_1}-\vec{r_2}|$ $\le$ 4 fm.
\cite{Puri96,Hart98}.\\
\begin{figure}[htbp]
\epsfxsize=8cm
\centerline{\epsfbox{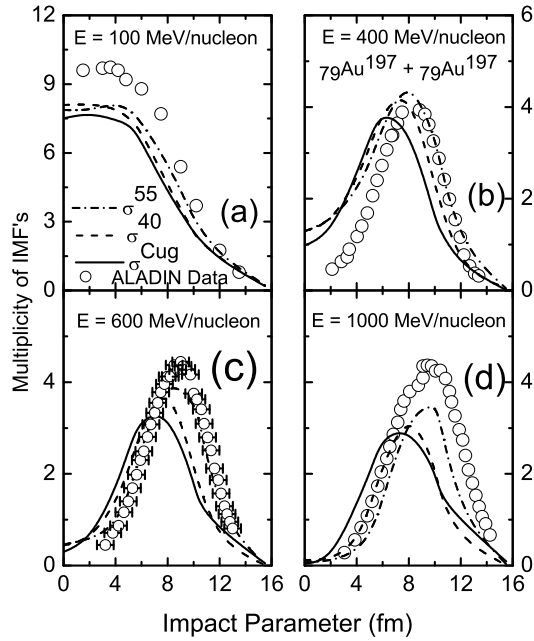}}
\caption{Comparison of average multiplicity of intermediate mass fragments (IMF's) with ALADIN data
at incident energies of 100, 400 MeV/nucleon (top panel) and 600, 1000 MeV/nucleon (bottom panel) as a function of impact parameter. The results are displayed using soft 
momentum dependent (SMD) interactions.}
\label{fig:1}
\end{figure}
In fig.\ref{fig:1},
we display the multiplicity of intermediate mass fragments (IMF's) as a function of
the impact parameter for the reaction of $_{79}Au^{197}~+~_{79}Au^{197}$ at
incident energies 100, 400, 600 and 1000 MeV/nucleon. Here soft momentum dependent interaction
(SMD) is
used with different cross-sections.
From the figure, the multiplicity of IMF's is maximal at 100 MeV/nucleon for smaller impact
parameters, which decreases with the increase in the impact parameter. 
On the other hand, one sees a rise and fall in the
multiplicity of the IMF's at higher incident energies. The dynamics at 100 MeV/nucleon, 
is mainly
governed by the mean field or by the density of the reaction as compared to other higher 
incident beam energies under consideration (e.g. 400, 600 and 1000 MeV/nucleon). The incident 
energy of 30 MeV/nucleon is the lowest limit for any semi-classical model, where the effect of 
Pauli-blocking is $\approx$ 90$\%$. Below this incident energy, quantum effects as well as pauli-
blocking need to be redefined. There are no visible
effects of different cross-sections at 100 MeV/nucleon. Due to the low
excitation energy, central collisions generate better repulsion and break the
colliding nuclei into IMF's, whereas for the peripheral
collisions, the size of the fragment is close to the size of the reacting nuclei, 
therefore, one sees a very few IMF's.
In contrary, a rise and fall can be seen at other higher incident energies. 
For the central collisions, the frequent NN collisions occurring at these energies do not allow any
IMF's production, whereas, at peripheral collisions the energy transfer is 
from the participating
matter to spectator matter is minimum, therefore, very few IMF's are seen.\\
In all the cases, some effects of different NN cross-sections are visible
at higher incident energies. The use of the momentum dependent interaction 
yields better comparison
with ALADIN setup \cite{Schu96} for $\sigma$ = 55 mb. This finding is 
in agreement with the results reported in ref.\cite{Puri96}, where it was found that NN
cross-section has sizable effect on reaction dynamics. Note that in these studies,
static equation of state was used.\\
\begin{figure}[htbp]
\epsfxsize=8cm
\centerline{\epsfbox{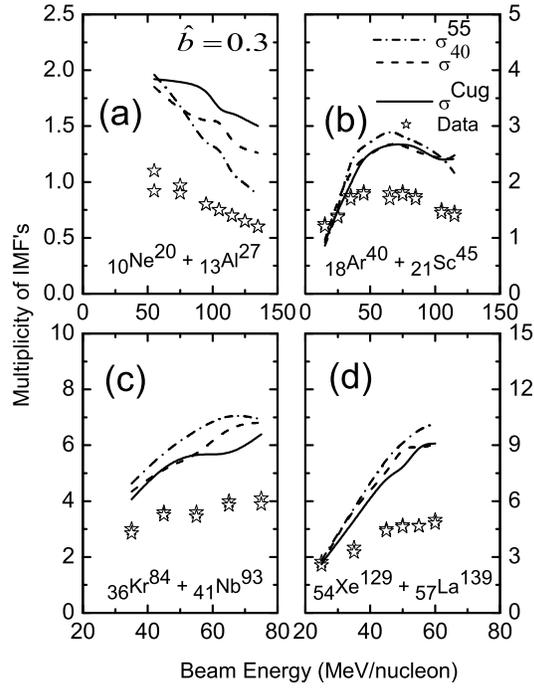}}
\caption{ The average IMF's multiplicity versus beam energy in the
reactions of $_{10}Ne^{20}~+~_{13}Al^{27},~
_{18}Ar^{40}~+~_{21}Sc^{45},~ _{36}Kr^{84}~+~_{41}Nb^{93}$ and $_{54}Xe^{129}~+~_{57}La^{139}$.
The symbols represent the NSCL experimental results,
while lines are
representing the results obtained within QMD model using different cross sections
$\sigma^{55}$, $\sigma^{40}$ and $\sigma^{Cug}$.}
\label{fig:2}
\end{figure}

Let us now analyze the above effects in asymmetric reactions .
In fig.~\ref{fig:2},
asymmetric reactions of $_{10}Ne^{20}~+~_{13}Al^{27}$, $_{18}Ar^{40}~+~_{21}Sc^{45}$, $_{36}Kr^{84}~+~_{41}Nb^{93}$ and $_{54}Xe^{131}~+~_{57}La^{139}$ are displayed as a function of beam energy at scaled impact parameter
$\hat{b} = 0.3$ (semi central collisions) using different cross-sections.
Final results are also compared with
the NSCL experimental data \cite{Llop95}. Due to no access to filters, no direct comparison with data could be made. These comparisons are just indicating the trend within theoretical framework. From the figure, it is clear that the trends of our
calculations with soft momentum dependent (SMD) interactions are in good agreement with experimental data. Again different NN cross-sections fail to make any significant impact on the cluster
dynamics.\\
One further see that the multiplicity of IMF’s in $_{10}Ne^{20} +_{13}Al^{27}$ decreases with
increase in the beam energies. For the reaction of
$_{18}Ar^{40}+_{21}Sc^{45}$, similar trends emerge above 55 MeV/nucleon, as in case of
$_{10}Ne^{20} +_{13}Al^{27}$. For energies below 55 MeV/nucleon, we see dominated role of
mean field and hence increase in the intermediate mass fragments.
For the rest of the
reactions, namely, $_{36}Kr^{84} +_{41}Nb^{93}$ and $_{54}Xe^{131} +_{57}La^{139}$,
the multiplicity of IMF's increases with the increase in the beam energy and
cross-section. One should note that in the first two reactions, incident energy is much
higher compared to the last two reactions.\\
\begin{figure}[htbp]
\epsfxsize=8cm
\centerline{\epsfbox{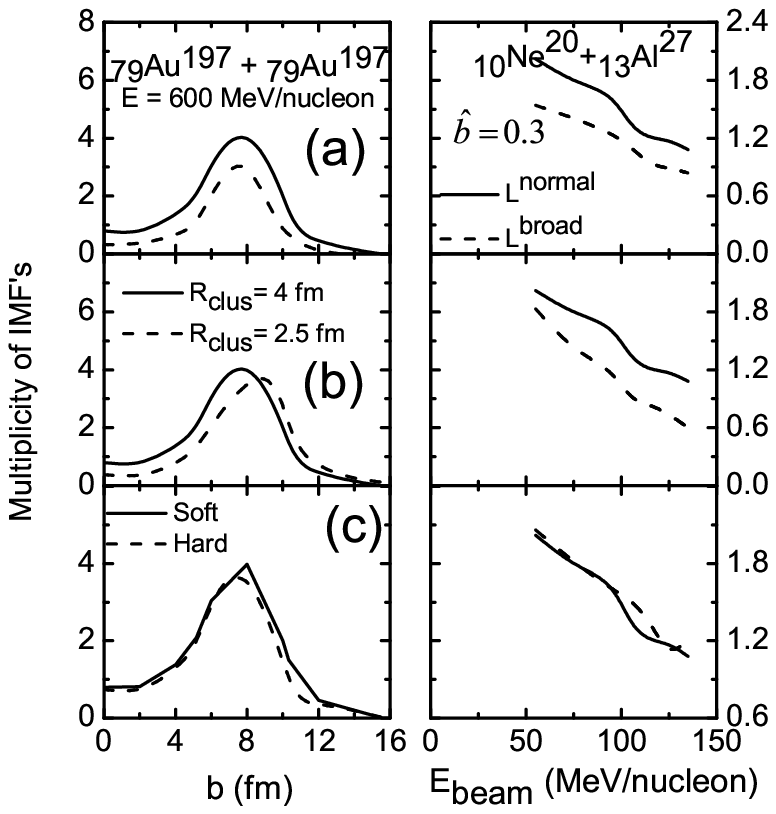}}
\caption{A Comparison of multiplicity of IMF's by employing different model ingredients like Gaussian width (upper panel),
clusterisation cut off distance $R_{Clus}$ (middle) and equation of state (bottom).
These results  are
displayed using  soft (upper, middle, bottom) and hard (bottom) equation of state
with $\sigma^{55}$.}
\label{fig:3}
\end{figure}

\begin{figure}[htbp]
\centerline{\epsfbox{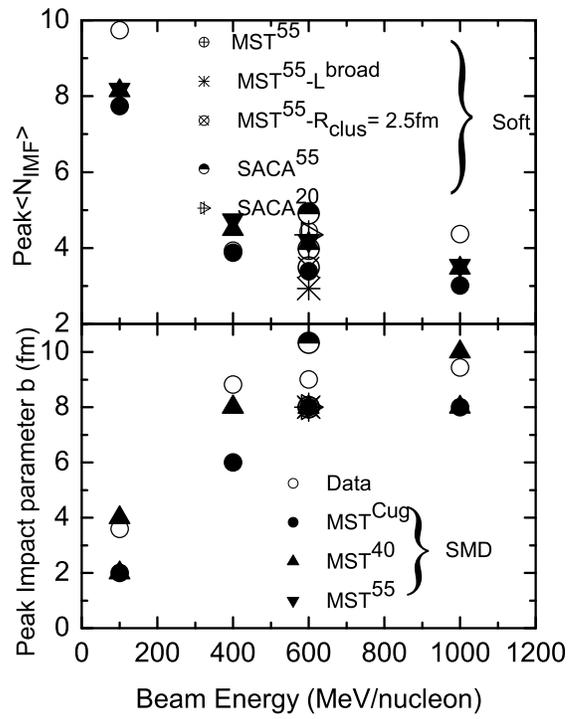}}
\caption{The Peak value of IMF’s and corresponding impact parameter as a function of
beam energy is compared with ALADIN data in the upper
and bottom panels, respectively.}
\label{fig:4}
\end{figure}

Let us now understand how other technical parameters affect the fragmentation. In fig.
~\ref{fig:3}, we
display in the upper part, the effect of width of Gaussian wave packet. We display the results with
narrow width ($L$ = 4.33 $fm^{2}$) and broad one ($L$ = 8.66 $fm^{2}$).  
We see that the variation of the width has
sizable effect on clusterisation. A broader Gaussian results in extended interaction
radius, therefore, binding more nucleons into a fragment. As a result, the fragment
turns much heavier than the upper limit of IMF's (i.e 65 for (Au+Au) reaction) and hence there
is net reduction in the production of IMF's. This result is in agreement with the 
findings in ref.\cite{Puri96}. One should also note that larger Gaussian width will also result
in more attractive nuclear flow and hence will push the energy of vanishing flow towards
higher incident energies.
In the middle panel, we display the effect of cut off distance $R_{Clus}$ on
clusterisation. We display the results with $R_{clus}~=~4$ fm and a narrower one 2.5 fm. We
see that when we choose a narrow cut off distance, lesser number of fragments are formed. In
the bottom panel, we display the role of different equations of state by simulating
the reactions with hard and soft equations of state. We see that different equations of state
do not alter the results.\\
To further analyze the results, we display in fig.~\ref{fig:4},
the maximal number of intermediate mass fragments
($N_{IMF}$) as well as corresponding impact parameter as a function of the 
incident energy for the
reaction of $_{79}Au^{197}~+~_{79}Au^{197}$. Here apart from the different Gaussian
widths and clusterisation distance, we also display the results with different 
clusterisation
algorithms. Here we also use Simulated Annealing Clusterisation Algorithm (SACA) 
\cite{Puri96}to
clusterise the phase space.
In SACA approach \cite{Puri96}, a group of nucleons can form a fragment if the total fragment energy/nucleon
$\zeta$ is below a minimum binding energy:
\begin{eqnarray} \label{ebind}
\zeta& =&  \sum_{i=1}^{N^f} \left [
\sqrt{({\bf p}_i - {\bf P}^{cm}_{N^f})^2 + m_i^2} - m_i +
\frac{1}{2} \sum_{j \ne i}^{N^{f}} V^{ij}
 ({\bf r}_i, {\bf r}_j) \right]\nonumber\\
&<& \mbox{$L_{be}$} \times N^f,
\end{eqnarray}
with $L_{be}$ =  -4.0 MeV if $N^f$ $\ge$ 3 and $L_{be}$ = 0 otherwise. In
this equation, $ N^f$ is the number of nucleons in a fragment,
${\bf P}_{N^f}^{cm}$ is the center-of-mass momentum of the fragment.
The requirement of a minimum binding energy excludes the loosly bound fragments which 
will decay
after a while.  To find the most bound configuration among the huge number of possible
fragmentation patterns, we proceed as follow: we start from a random configuration
which is chosen by dividing the whole system into few fragments. The energy of the individual clusters is calculated by summing over all nucleons present in that cluster using eqn.\ref{ebind}.
Note that as we neglect the interaction between fragments, the total energy calculated in this
way differs from the total energy of the system. If the difference between the old and 
new energy is
negative, the new configuration is accepted. If not, the new configuration may
nevertheless be accepted with a probability of exp(-$\Delta$E/c), where c is
the control parameter. This procedure is known as the Metropolis algorithm. The
control parameter is decreased in small steps. This algorithm will
yield eventually the most bound configuration. The identification of fragments using SACA is
very time consuming. The algorithm searches the most bound configuration out of several
million of different possibilities.  The major outcome of this method is that we can
recognize the fragments at relative high density.\\

From the figure, we see that different cross-sections yield
quite similar trend. The absolute value varies with NN cross-sections as
well as with other model ingredients. For example, we see that at 600 MeV/nucleon, the 
SACA method with
20 mb yields results close to the MST method with 55 mb. A broader Gaussian scales down the
number and so
is the case with reduced cut off distance. From this discussion, it is clear that 
the effect of larger
cross-section (i.e. between 20 and 55 mb) is as large as effect of other model ingredients.
Even a change of the clusterisation algorithm affect the outcome sizeably.
From this analysis,
one conclude that it may not be possible to pin down the magnitude of cross-section from
multifragmentation since other technical parameters such as
width of the Gaussian, clusterisation range or even the change of the
clusterisation algorithm
alters the results in similar fashion when semi-classical model such as quantum 
molecular dynamics is used.\\

\section{Summary}
In the present study, we focused on the comparative study of different
model ingredients with semi-classical model, namely, quantum molecular
dynamics (QMD) model. For this study, a hunt was made to
compare our theoretical results of different cross-sections with experimental data. We also
analyzed the results with model ingredients such as width of the Gaussian, clusterisation
range and different clusterisation algorithms. We found that the effect of different
cross-sections is of the order of the one obtained from the model ingredients. All model
ingredients affect the fragmentation pattern in a similar fashion.\\

\section{Acknowledgment}
This work has been supported by the Grant no. 03(1062)06/ EMR-II, from the Council of Scientific and
Industrial Research (CSIR) New Delhi, Govt. of India.

\end{document}